\documentclass[%
 reprint,
 superscriptaddress,
 aps,
 prl,
]{revtex4-1}

\usepackage{amsmath,amssymb}
\usepackage{graphicx}
\usepackage{nicefrac}
\usepackage{newtxtext,newtxmath}
\usepackage{microtype}

\newcommand{\beq}{\begin{equation}}
\newcommand{\eeq}{\end{equation}}

\begin{document}

\title{BKT-paired phase in coupled XY models}

\author{Giacomo Bighin}
\thanks{These two authors contributed equally.}
\affiliation{IST Austria (Institute of Science and Technology Austria), 
Am Campus 1, 3400 Klosterneuburg, Austria}

\author{Nicol\`o Defenu}
\thanks{These two authors contributed equally.}
\affiliation{Institut f\"ur Theoretische Physik, Universit\"at Heidelberg, D-69120 Heidelberg, Germany}

\author{Istv\'an N\'andori}
\affiliation{MTA-DE Particle Physics Research Group, P.O.Box 51, H-4001 Debrecen, Hungary}
\affiliation{MTA Atomki, P.O.Box 51, H-4001 Debrecen, Hungary} 
\affiliation{University of Debrecen, P.O.Box 105, H-4010 Debrecen, Hungary}

\author{Luca Salasnich}
\affiliation{Dipartimento di Fisica e Astronomia ``Galileo Galilei'',
Universit\`a di Padova, Via Marzolo 8, 35131 Padova, Italy}
\affiliation{Istituto Nazionale di Ottica (INO) del Consiglio Nazionale
delle Ricerche (CNR), Sezione di Sesto Fiorentino, 
Via Nello Carrara 2, 50019 Sesto Fiorentino, Italy}

\author{Andrea Trombettoni}
\affiliation{CNR-IOM DEMOCRITOS Simulation Center, Via Bonomea 265, I-34136 Trieste, Italy}
\affiliation{SISSA and INFN, Sezione di Trieste, Via Bonomea 265, I-34136 Trieste, Italy}

\date{\today}

\begin{abstract}
We study the effect of a linear tunneling coupling between 
2D systems, each separately exhibiting the topological 
Berezinskii-Kosterlitz-Thouless (BKT) transition. In the uncoupled limit, 
there are two phases: one where
the $1$-body correlation functions are algebraically decaying and the other 
with exponential decay. When the linear coupling is turned on,
a third BKT-paired phase emerges, in which $1$-body correlations are
exponentially decaying, while $2$-body correlation functions exhibit power-law decay. 
We perform numerical simulations in the paradigmatic case of two coupled 
$XY$ models at finite temperature,
finding that for any finite value 
of the interlayer coupling, the BKT-paired phase is present. 
We provide a picture of the phase diagram using a 
renormalization group approach. 
\end{abstract}

\maketitle

An ample variety of physical properties and phenomena 
emerge when two many-body systems are coupled. There are, of course, 
different ways of coupling interacting systems, depending on the geometry 
of the uncoupled systems, on the relevant degrees of freedom and on the 
way in which the coupling is established. 
One of them is the so-called {\em weak} coupling, when the Hamiltonian 
term modeling the coupling is a perturbation with respect to the 
uncoupled Hamiltonians. In this case, if the two systems are 
separately described by a macroscopic wavefunction -- such as superfluids or superconductors --
a relative phase between them emerges,
as one can see in the Josephson effect between 
superconductors \cite{Barone82,Tafuri05}, 
Helium containers \cite{Pereverzev97,Hoskinson05}, Bose-Einstein condensates 
\cite{Cataliotti01,Albiez05} and ultracold fermions \cite{Valtolina15}. 
In these cases, the microscopic details of the coupling enter only in the 
Josephson energy ruling the maximum current that can flow through the junction.

Coupling two superfluids gives rise to a dissipationless drag current that is often referred to as the Andreev-Bashkin effect\,\cite{Andreev76}. Previous investigations\,\cite{Kaurov05,Kuklov03,Svistunov15} have mainly focused on the three-dimensional case, where strong enough drag densities have been found to modify the order of the symmetry breaking transition\,\cite{Dahl08}. Indeed, when the coupling is no longer a perturbation, one can expect 
that the bulk properties of the uncoupled systems are significantly 
altered and that new phases may emerge. A typical phenomenon 
induced by a strong 
coupling is that the 
order {\em in} the systems is substituted 
by a phase {\em between} them, therefore giving rise to an 
order parameter expressed in terms of operators of both systems. 
Countless different examples of couplings 
exist. Among the simplest instances, 
one can imagine taking two one-dimensional systems 
and couple them lengthwise, thus creating a ladder 
geometry. Examples of this kind of coupling include magnetic spin ladders, 
where two quantum spin chains are put in a ladder geometry, see for instance Refs.~\cite{Gogolin98,Dagotto99}. The bilayer structure, 
in which two two-dimensional 
systems are coupled, is also a paradigmatic configuration,
which has been studied in numerous 
physical systems, ranging from graphene 
\cite{Novoselov04} to quantum Hall systems 
\cite{Jain07} and dipolar models \cite{Baranov12}. The bilayer 
configuration is of particular relevance because it may 
refer to two quantum or classical systems at finite temperature, or -- 
via the quantum-classical correspondence -- 
to the ladder geometry in which two quantum one-dimensional 
models at zero temperature are coupled.

In this Letter, we address the emergence of a paired phase when two models 
exhibiting a topological phase transition are coupled. We will 
refer to the Berezinskii-Kosterlitz-Thouless (BKT) universality 
class \cite{Kosterlitz16} 
and study the effect of the coupling of two BKT systems. We have two main motivations for 
our study. First, the BKT transition is not characterized by a local order 
parameter and by the conventional spontaneous symmetry breaking. Rather, 
one can locate the critical point in which the system becomes superfluid by 
looking at the correlation functions, exhibiting 
a power law (exponential) decay in (outside) the superfluid phase. BKT transitions 
has been observed in several different 2D physical systems, 
including superfluid Helium \cite{Bishop78}, superconducting films and arrays 
\cite{Simanek1994,Fazio94}, as well as bosonic \cite{Hadzibabic2006,Schweikhard07} 
and fermionic \cite{Murthy15} ultracold systems.

The second motivation is related to the presence and relevance of the 
coupling between 2D systems 
in a variety of experimental systems, 
ranging from layered superconductors \cite{Tinkham96,Klemm12} 
2D two-component mixtures \cite{Svistunov15} and 2D ultracold gases \cite{Hadzibabic11}. 
In the latter case, a single 2D ultracold system is obtained 
by means of a suitably large transverse confinement: for instance, one can realize a 2D geometry 
by confining the atoms in a well of an optical lattice along the transverse direction, with the 
transverse confinement frequency increasing with the power laser and being $\gtrsim 1-10 \text{ kHz}$ \cite{Bloch08}. 
When two adjacent minima of the vertical lattice are left populated, one would have 
two 2D ultracold gases with tunable interlayer coupling. In this way, one would study the effect 
of the bilayer coupling on the 2D physics already observed in single layers 
\cite{Hadzibabic2006,Schweikhard07,Murthy15}. Notice that in this case, 
the coupling is (for bosons) of the form $b^\dag_1 b_2$ 
(where $b_\alpha$ creates a particle in the $\alpha$ layer), thereby
corresponding to what we refer to as linear coupling. 

Despite the rising interest of the theoretical community in low dimensional binary mixtures \cite{Karle19,Parisi18,Konietin18,Sellin18}, the influence of interspecies coupling on universal behavior remains largely unknown. In this context, a compelling question is {\em i)} ``What is the fate of the quasi-long-range order in the presence of a strong coupling between two BKT systems?''. One may as well ask {\em ii)} ``Does quasi order occur in the mixed correlators of the coupled system? If so, does it appear already at small couplings?". Previous attempts to answer these questions mainly focused on the study of effective models for coupled topological excitations, such as the coupled sine-Gordon model\,\cite{Mathey08} and the coupled Coulomb gases\,\cite{Podolsky09}, evidencing the emergence of a composite vortex gas phase as well as a second-order phase transition at strong coupling. However, the results of these investigations cannot be directly applied to our case, since the low energy equivalence between two dimensional superfluids and topological defect gases appears to be spoiled by the presence of the bilayer coupling term.

To address the questions above, we consider the paradigmatic case of two coupled $XY$ models at finite temperature, based on the expectation that similar features emerge for other classical and quantum models in the $XY$ universality class when linearly coupled. The paradigmatic nature of the XY model results in longly certified numerical tools, which allow us to reliably investigate the bilayer case.

\paragraph{\textbf{The model.}} The bilayer $XY$ Hamiltonian can be written as 
\begin{multline}
\mathcal{H} = - J \sum_{\langle i j \rangle} \cos \left( \phi_i - \phi_j \right) - 
J \sum_{\langle i j \rangle} \cos \left( \psi_i - \psi_j \right) + \\
- K \sum_{i} \cos \left( \phi_i - \psi_i \right), 
\label{eq:halt} 
\end{multline}
where $\vec{S}_i = (\cos \phi_i, \sin \phi_i)$ and $\vec{T}_i = (\cos \psi_i, \sin \psi_i)$ are the $XY$ spins with $n=2$ components defined on the first and the second layer, respectively, each being a 2D system. We will also use the notation $S_i=e^{\mathrm{i}\phi_i}$ 
and $T_i=e^{\mathrm{i} \psi_i}$. Notice that we are considering the symmetric case 
where the intralayer coupling takes the same value $J$ on both layers. The critical line separating 
the region with power law correlations from the one 
with exponential correlations in model (\ref{eq:halt}) has been studied in 
literature \cite{parga}. Moreover, several properties of coupled and 
layered XY models and two-components systems have also been investigated
\cite{Granato88,Zhang05,Vayl17,Rancon17,Kobayashi18,Karle19}, as well as 
coupled $3D$ $XY$ models \cite{Smiseth05,Sellin16}. Here our focus 
is on the remarkable quasi-order for pairs of spins on different layers that arises in a 2D bilayer system.

When $K=0$, the two layers are decoupled and the behavior of the model 
is that of the standard 2D XY model 
\cite{Kosterlitz16,Svistunov15}: a superfluid phase and the normal state are separated by the BKT transition occurring at $T_\text{BKT} \approx 0.893 J$ 
\cite{Gupta1988,Gupta92,tomita,Hasenbusch05,Komura12}. 
The normal state is characterized by the exponential decay of the phase-phase correlator; in the superfluid phase, on the other hand, one finds the peculiar algebraic decay of correlations functions.

In the other limiting case, $K \to \infty$, 
the spins on the upper and lower layer are constrained to be parallel, 
{\em i.e.}, $\phi_i = \psi_i \ \forall i$, so that the 
effective Hamiltonian reduces, up to additive constants, to
\beq
\mathcal{H} = - 2J \sum_{\langle i j \rangle} \cos \left( \phi_i - \phi_j \right),
\label{eq:hlimit} 
\eeq
{\em i.e.}~that of a 2D $XY$ model with coupling constant $J'=2J$, 
therefore with critical temperature $T'_{BKT} \approx 1.786 \ J$.

\paragraph{\textbf{Results for the phase diagram.}} We want to study the full phase diagram of the model, 
away from the two limiting cases just analyzed. From the theory  
of bilayer systems and two-component mixtures \cite{Svistunov15,Ueda10} one may 
expect that, apart from the normal and superfluid phases of the BKT transition, 
a phase involving variables (operators in the quantum case) of both layers 
can emerge. The question is whether it emerges only for large values of the 
coupling $K$ or not. To make a comparison with a similar, yet different, 
analytically solvable case, let us consider the square lattice Askhin-Teller model 
\cite{baxter}: the Hamiltonian has a form similar to Eq. (\ref{eq:halt}), 
$H_{AT}=-J \sum_{\langle i j \rangle} s_i s_j -J \sum_{\langle i j \rangle} t_i t_j 
-K \sum_{\langle i j \rangle} s_i s_j t_i t_j$, with $s_i, t_i=\pm 1$. 
The model has indeed a phase 
whose order parameter is $\langle s t \rangle$ \cite{baxter}. 
However, apart from the fact 
that spin variables are discrete in the Askhin-Teller model and continuous 
in the bilayer $XY$ model, there are two remarkable differences: first, 
the coupling 
is quartic; second, the bilayer $XY$ model exhibits quasi-order and no 
spontaneous symmetry braking at finite temperature.

To determine the full phase diagram of the bilayer $XY$ model,
we perform Monte Carlo (MC) simulations of the 
Hamiltonian of Eq.~(\ref{eq:halt}) on a $2 \times L \times L$ lattice, 
$L$ being the linear dimension of each layer. We vary 
$\beta J$ and $\beta K$, where as usual $\beta=(k_B T)^{-1}$. 
Monte Carlo updates use the Swendsen-Wang algorithm \cite{tomita,swendsen} 
after embedded cluster decomposition \cite{wolff}. 
We sample the following correlation functions:
\begin{align}
c_\uparrow (k) &= \sum_{|i - j| = k} \exp(\mathrm{i} \phi_i - \mathrm{i} \phi_j) \\
c_\downarrow (k) &= \sum_{|i - j| = k} \exp(\mathrm{i} \psi_i - \mathrm{i} \psi_j) \\
z (k) &= \sum_{|i - j| = k} \exp(\mathrm{i} \phi_i + \mathrm{i} \psi_i - \mathrm{i} \phi_j - \mathrm{i} \psi_j) 
\end{align}
where the summations extend over all $i$, $j$ pairs separated by $k$ lattice sites along the $x$ or $y$ direction. Due to the symmetry of the system upon exchange between the upper and lower layer, $c_\uparrow (k)$ and $c_\downarrow (k)$ coincide across the whole phase diagram, let us call them $c(k)$. All correlation functions are well fitted by the function $Q(r) = ( r / r_\text{alg} )^{\alpha} \exp (- r/r_\text{exp})$, 
where $r_\text{alg}$ is the characteristic length associated to the algebraic decay and $r_\text{exp}$ is the characteristic length associated to the exponential decay.

\begin{figure*}
\centering
\includegraphics[width=1.00\linewidth,trim={8pt 0 8pt 0},clip]{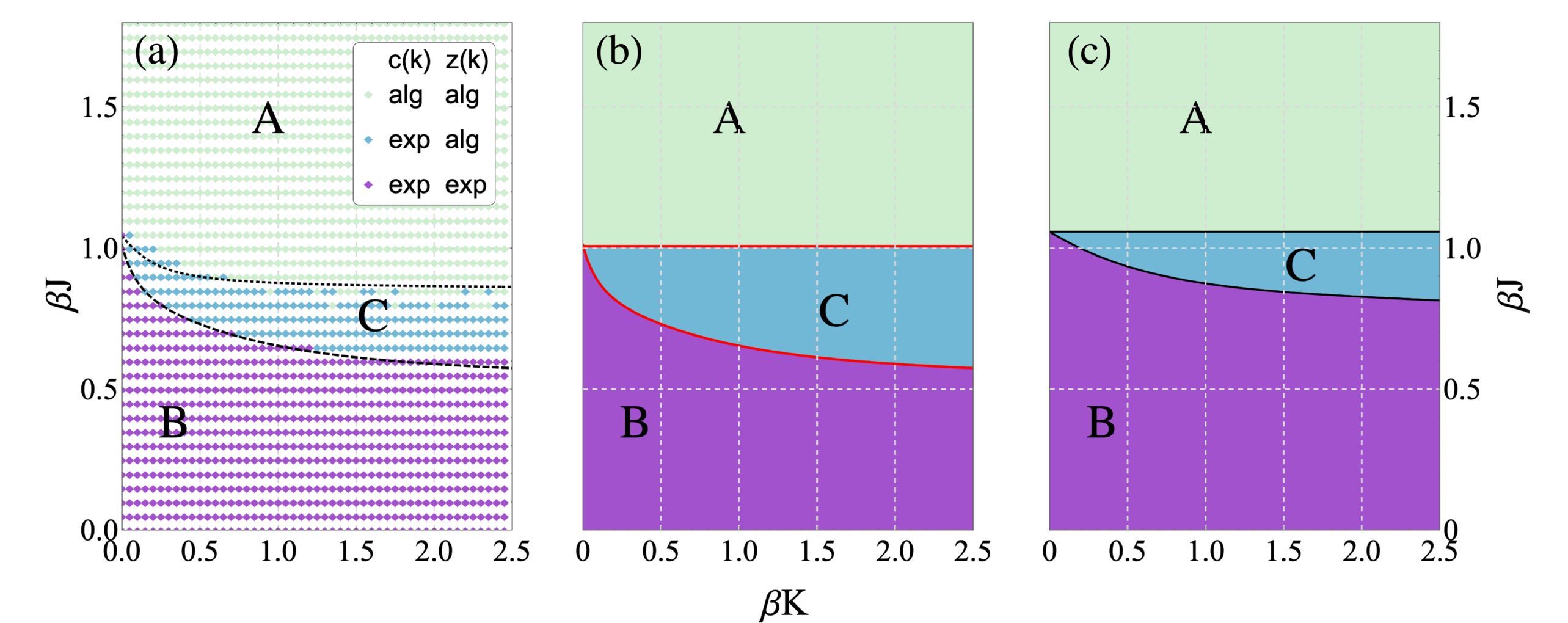}
\caption{$\beta J$-$\beta K$ phase diagram, showing the paired phase $C$ 
appearing  for finite values of $\beta K$. Left panel: Monte Carlo results 
($L=64$) obtained from the fit of correlation functions $c(k)$ and $z(k)$ as 
described in the text. The lower line has been obtained using the PCC technique, as in the next panel, while the upper line guides the eye. Middle panel: Monte Carlo results (L=64) obtained through the
PCC technique, allowing the Kasteleyn-Fortuin clusters to extend
on a single layer only (upper red line) or on both layers (lower red line).
Right panel: RG phase diagram, see main text.}
\label{fig:phasediagram}
\end{figure*}

To tell apart the exponential and algebraic behaviour 
we compare $r_\text{exp}$ with the linear system size used in the MC 
simulation: if $r_\text{exp}$ is smaller than twice the lattice dimensions, 
then we mark the decay as exponential, otherwise it is marked as algebraic, 
since the exponential characteristic length $r_\text{exp}$ is 
neglectable when compared to the dimensions of the system being studied. 
This allows us to obtain the phase diagram shown in 
Fig. \ref{fig:phasediagram}(a), 
identifying the three phases. We denote by $A$ the usual superfluid BKT phase, 
characterized by the algebraic decay of $c(k)$. The correlation function $z(k)$ is also found to be algebraically decaying in 
the $A$ phase.
The normal phase, denoted by $B$, is characterized, as usual, by the exponential decay 
of $c(k)$. We checked that in the $B$ phase $z(k)$ is exponentially
decaying, as well. Finally, we observe a third $C$ phase, that is instead characterized by the exponential 
decay of $c(k)$ and the algebraic decay of $z(k)$. Notice that, 
within the precision we can achieve, for small values of $K$ one always has a range
of values of $J$ for which there is the $C$ phase. In other words, the $C$ phase 
extends up to vanishing values of $K$.

We dub the $C$ phase {\em `BKT-paired phase'}, since each layer
has the $1$-body correlation function $c(k)$ decaying exponentially, 
while the $2$-body correlation function $z(k)$ has an algebraic behaviour. 
Since $z(k)$ is the correlation function between $S_i T_i$ and 
$S_j T_j$ for $i$ and $j$ far apart from each other, this is related to a 
quasi-long-range order for the pair variables $S_i T_i$ 
in the different sites of the bilayer system. 
The phase transition between $A$ and $C$ appears to be a conventional 
BKT line, and in agreement with Ref.~\cite{parga}. It is worth noting that
despite some similarities between the phase diagram presented in Fig.\,\ref{fig:phasediagram}
and the one discussed in Ref.\,\cite{Mathey08}, we find no trace of any actual symmetry breaking while
our paired BKT phase appears also at infinitely small coupling strengths. Moreover, the presence
of power law four point correlations was not noticed in previous investigations.

To further numerically investigate these two transition lines 
we resort to probability-changing cluster (PCC) algorithm, introduced in Ref.~\cite{tomita}
for the single-layer BKT transition. Following the approach therein,
at first the XY variables are mapped onto two Ising models, using
Wolff's embedded cluster formalism \cite{wolff}. Subsequently the temperature of the
system is tuned in progressively smaller increments, looking for the $\nicefrac{1}{2}$
percolation threshold of the Kasteleyn-Fortuin clusters defined from each
Ising model \cite{tomita}, signaling the BKT transition. It can in fact be shown \cite{coniglio,kf}
that the spin-spin correlation function $\langle S_i S_j \rangle$ equals $p_{ij}$,
the probability that the site $i$ and $j$ belong to the Kasteleyn-Fortuin same cluster, effectively linking
the presence of a percolating cluster spanning the whole system to the
onset of quasi-long-range order. In the present case one
can identify two different critical lines through the
$\nicefrac{1}{2}$ percolation threshold criterion,
depending on whether the clusters by which percolation is identified
are allowed to extend on a single layer or on both layers, respectively.
The results we obtain, shown in Fig. \ref{fig:phasediagram}(b), are compatible
with the  identification of the $A$, $B$ and $C$ phases obtained by the study 
of correlation functions. We now supplement this analysis by a RG 
calculation.

\paragraph{\textbf{RG approach.}} Mermin-Wagner theorem\,\cite{Mermin1966} forbids spontaneous breaking of continuous symmetries in two dimensions and, thus, only quasi-order, characterised by power law correlations, can arise in our model. Such effect, which cannot be reproduced within traditional perturbative approaches, can only be described by RG theory via the explicit introduction of the topological defects\,\cite{Jose1977, Kosterlitz16}. Within the functional RG framework\,\cite{Delamotte2012,Wetterich1993},  it has been possible to derive non-perturbative flow equations which correctly reproduce the Mermin-Wagner theorem\,\cite{Codello2013,Codello2015,Defenu2015} for $O(N)$ symmetric model, but the presence of quasi-order in the thermodynamic limit for $N=2$ and $d=2$ could not be found\,\cite{Grater1995}. Recently, a new method has been proposed\,\cite{Defenu2017} which allows to include in the traditional BKT formalism the non-universal corrections arising at small and intermediate scales.

We shall now employ a simplified version of this approach to the bilayer case, showing how it is possible to reproduce the MC phase diagram. First of all, we are going to derive the mean field solution of our problem; then we are going to use it to compute the effective spin stiffness in the low temperature phase,  which will serve as the initial condition of the traditional BKT flow equations\,\cite{Kosterlitz1973,Kosterlitz1974}. Finally, solving the BKT flow equations with the MF initial condition, we will be able to locate the vortex unbinding transition.

Within our framework it is convenient to employ a Hubbard-Stratonovich transformation\,\cite{Simanek1994, Machado2010, Nishimori2011}, in order to introduce the continuous field representation of the bilayer model under study. Following this procedure, the bilayer XY model can be exactly mapped into two coupled complex $|\varphi|^{4}$ theories with the action
\begin{align}
\label{mf1}
S[\varphi]=S_{\mathrm{kin}}[\varphi]+\sum_{l}\int{U(|\varphi_{l}|)d^{2}x}
\end{align}
where $U(|\varphi|)=\log(I_{0}(|\varphi|))$ is the local potential of the model, written in terms of the zeroth order Bessel function $I_{0}$, and $S_{\mathrm{kin}}[\varphi]$ is the kinetic action. The index $l\in\{1,2\}$ label two different complex fields, which represent the spin variables in the two different layers.

The kinetic part of the action is readily written in Fourier space
\begin{align}
\label{mf2}
S_\text{kin}[\phi]=\frac{1}{2}\sum_{\sigma,q}\frac{\varphi_{\sigma}(q)\varphi_{\sigma}(q)}{K^{\sigma}(q)},
\end{align}
where $K^{\sigma}(q)=2J\varepsilon_{0}(q)+2\mu+\sigma 2K$, $\sigma\in\{+,-\}$ and a multiplicative $\beta$ factor has been absorbed into the definition of the couplings. The coefficient $\mu$ has to be introduced in order to have a positively defined exchange matrix, as needed in the Hubbard-Stratonovich transformation\, \cite{Machado2010}. The relation between the $\varphi_{\pm}$ fields and the original $\varphi_{1,2}$ counterparts reads
\beq
\label{mf3}
\varphi_{\pm}(q)=\frac{\varphi_{1}(q)\pm\varphi_{2}(q)}{\sqrt{2}} \; .
\eeq

Rewriting the local potential in Eq.\,\eqref{mf1} in terms of the $\varphi_{\pm}$ fields just introduced, one can identify the masses
\beq
\label{mf4}
\mu_{\pm}=\left(\frac{1}{K^{\pm}(0)}-\frac{1}{2}\right) \; .
\eeq
In order to proceed further with the analysis, one has to choose a suitable value of $\mu$. In principle any value of $\mu>2J+K$ is a valid choice and the choice of $\mu$ shall not alter the behavior of the model\,\cite{Machado2010}. However, when approximations are employed, a $\mu$-dependence of the result can be found\,\cite{Jakubczyk2016,Defenu2017}. In the following, we are going to only consider the continuum limit of the kinetic action in Eq.\,\eqref{mf2}, discarding lattice effects, thus we may extend our choices also to $\mu<2J+K$ and, in particular, we will fix $\mu=K$, consistently with the optimal choice discussed in\,\cite{Defenu2017} for a single XY layer.

It is now possible to compute the two MF phase boundaries imposing the vanishing of one of the two masses $\mu_{\pm}$ at each phase boundary. According to this argument the lower phase boundary in Fig. \ref{fig:phasediagram}(c) is obtained by the vanishing of $\mu_{+}$, while $\mu_{-}$ becomes zero at the upper boundary. Above each boundary one of the fields $\varphi_{\pm}$ acquires a finite expectation value thus violating the Mermin-Wagner theorem.
The field expectation values can be obtained solving the saddle point equation
\begin{align}
\label{mf5}
0=\frac{\partial S[\varphi]}{\partial\varphi_{\pm}}\bigg{|}_{\varphi=\text{const}}
\end{align}
where $S[\varphi]$ is defined in Eq.\,\eqref{mf1}. The field density is defined according to $\rho_{\text{MF},\sigma}=|\varphi_{\text{MF},\sigma}|^{2}$, with $\varphi_{\text{MF},\sigma}$ solution of the saddle point equation, see Eq.\,\eqref{mf5}.

As anticipated, the low energy behaviour of the model cannot be described simply considering the static MF solution, which provides an unphysical symmetry breaking scenario. In order to correct the MF picture and provide a reliable phase diagram, the effect of vortex fluctuations at low energy shall be included. In this perspective, it is convenient to consider the action\,\eqref{mf1} in the continuum limit
\begin{align}
S_\text{c}[\varphi]=\sum_{\sigma}\int\,d^{2}x\frac{|\nabla\varphi_{\sigma}(x)|^{2}}{2m_{\sigma}}+\text{local terms}
\end{align} 
where $m_{\pm}^{-1}=\frac{J}{K^{\pm}(0)^{2}}$. The explicit introduction of vortex fluctuations shall be done in the amplitude and phase representation $\varphi_{\sigma}=\sqrt{\rho_{\sigma}}e^{\mathrm{i}\theta_{\sigma}(x)}$. Assuming that the field amplitude remains frozen at the MF expectation $\rho_{\sigma}=\rho_{\text{MF},\sigma}$, the model action only contains phase terms
\begin{align}
\label{mf7}
S[\theta]=\sum_{\sigma}\int\,d^{2}x\frac{\rho_{\sigma}}{2m_{\sigma}}\partial_{\mu}\theta_{\sigma}\partial_{\mu}\theta_{\sigma}.
\end{align}
In the latter action the two phases are uncoupled and two distinguished BKT transitions  are found. Latter statement can be proven by mapping each independent quadratic phase term into a sine-Gordon model\,\cite{Defenu2017}.

The physics of the transitions is fully described, at this approximation level, by the effective phase stiffness $\rho_{\sigma}/2m_{\sigma}$  which is non vanishing only for finite density expectation $\rho_{\sigma}>0$. Once again for fixed $K$ and $J<1-K+\mu$ both the effective phase stiffnesses vanish and all the system correlation functions are exponentially decaying. Then for $1-K+\mu<J<1+K+\mu$ the effective stiffness of the $\theta_{+}$ phase becomes finite and the phase dynamics becomes relevant. The low energy phase dynamics is conveniently described by the BKT flow equations
\begin{align}
\partial_{t}K_{k}&=-\pi g_k^{2}K_k^{2},\label{Eq23}\\
\partial_{t}g_{k}&=\pi\left(\frac{2}{\pi}-K_k\right)g_k\label{Eq24}
\end{align}
At the bare level, $K$ and $g$ assume the values
\begin{align}
K_{\Lambda}&=J_{\mathrm{eff}}^{\sigma},\label{Eq25}\\
g_{\Lambda}&=2\pi e^{-\pi^{2}K_{\Lambda}/2}\label{Eq26}
\end{align}
where the bare effective spin system is simply given by
\begin{align}
\label{eff_stiff}
J_{\mathrm{eff}}^{\sigma}=\frac{\rho_{\text{MF},\sigma}}{2m_{\sigma}}.
\end{align} 
In order to derive Eqs.~\eqref{Eq23}--\eqref{Eq24}
one has to assume a UV regularization, which traditionally relies in
considering the Coulomb gas charges as hard disks of finite
radius \cite{Kosterlitz1973}.

Using the MF expectations obtained solving the saddle point Eq.\,\eqref{mf5} we can now derive the theoretical prediction for the two BKT transition line reported in panel (b) of Fig. \ref{fig:phasediagram}. The upper transition line, which corresponds to the conventional BKT transition, comes from the unbinding of the $\theta_{-}$ vortexes. Such phase boundary appears to be flat at our approximation level since we solve Eq.\,\eqref{mf5} separately for $\varphi_{\text{MF}, -}$ posing  $\varphi_{\text{MF},+}=0$, due to the difficulties encountered in solving the two coupled equations for the expectation values. On the other hand the $K$ dependence of the lower transition line is only due to the explicit $K$ dependence in the $m_{+}$ factor and it is in good agreement with the MC prediction. 

Finally, we should compare our results with the corresponding
  results for two linearly-coupled Ising models. Denoting by $S_i$ and $T_i$
  the spins in the two layers, a phase
  with order parameter $\langle S T \rangle \neq 0$ cannot exist
  (in the paramagnetic phase with
  $\langle S \rangle = \langle T \rangle = 0$).
  The same is occuring for the two linear coupled $XY$
  models treated in this Letter, {\it i.e.}~
  there is no phase with order parameter
  $\langle \cos(\phi+\psi)\rangle$,
  in agreement with the Mermin-Wagner theorem.
  What is present is indeed a BKT of the pairs, specific
  of coupled {\it two-dimensional} systems with $O(2)$ symmetry, 
  not contradicting the previous result.

\paragraph{\textbf{Conclusions.}}  We investigated the occurrence of a BKT-paired phase in coupled 
$XY$ models, corresponding to an exponential (algebraic) 
decay for the $1$-body ($2$-body) correlation functions. We found that 
the third phase is present even for very small linear couplings 
between the layers, 
within the precision of our simulations. Furthermore, we presented an RG treatment 
which was found to be in agreement with the aforementioned results.
Our findings call for a more systematic study of finite size-scaling effects 
in the $J-K$ phase diagram: experiments with coupled 2D ultracold gases are 
performed on lattices with a linear size of tens up to a few hundreds of lattices 
sites, therefore motivating the study of the presence 
of the BKT-paired phase in realistic setups. 

Furthermore, an exciting perspective is represented by a comparison of the present results
with the ones obtained for other types of couplings, such as quartic ones, and
to consider the case of different $J$'s in the layer. 
A systematic analysis for small $K$ done at larger sizes would also be 
important in clarifying
whether the critical value for the presence of the BKT transition is zero 
as the RG approach seems to suggest. Moreover, we think as well that it would be a 
deserving subject of future investigations to further analyse
the order to quasi-order transition between phases $B$ and $C$ of 
Fig.~\ref{fig:phasediagram}.
Finally, we mention that an interesting perspective is represented by the comparison of our results with analytical findings
  from suitably derived theories where two sine-Gordon models are coupled
in a way fixed by the microscopic coupling.\\

\begin{acknowledgments}
We thank S.~Chiacchiera, G.~Delfino, N.~Dupuis, T.~Enss, M.~Fabrizio and G.~Gori for many stimulating discussions. G.B.~acknowledges support from the Austrian Science Fund (FWF), under project No. M2461-N27. N.D.~acknowledges support from Deutsche Forschungsgemeinschaft (DFG) under Germany's Excellence Strategy EXC-2181/1 - 390900948 (the Heidelberg STRUCTURES Excellence Cluster) and from the DFG Collaborative Research Centre ``SFB 1225 ISOQUANT''. Support from the CNR/MTA Italy-Hungary 2019-2021 Joint Project \textit{``Strongly interacting systems in confined geometries''} is gratefully acknowledged. 
\end{acknowledgments}

\end{document}